# Predicting Glass-to-Glass and Liquid-to-Liquid Phase Transitions in Supercooled Water using Classical Nucleation Theory

**Robert F. Tournier[1,2]**

[1]*Univ. Grenoble Alpes, Inst. NEEL, F-38042 Grenoble Cedex 9, France*

[2]*CNRS, Inst. NEEL, F-38042 Grenoble, France*

*E-mail address: robert.tournier@neel.cnrs.fr*

**Abstract:**

Glass-to-glass and liquid-to-liquid phase transitions are observed in bulk and confined water, with or without applied pressure. They result from the competition of two liquid phases separated by an enthalpy difference depending on temperature. The classical nucleation equation of these phases is completed by this quantity existing at all temperatures, a pressure contribution, and an enthalpy excess. This equation leads to two homogeneous nucleation temperatures in each liquid phase; the first one ($T_{n-}$ below $T_m$) being the formation temperature of an "ordered" liquid phase and the second one corresponding to the overheating temperature ($T_{n+}$ above $T_m$). Thermodynamic properties, double glass transition temperatures, sharp enthalpy and volume changes are predicted in agreement with experimental results. The first-order transition line at $T_{LL}$=0.833×$T_m$ between fragile and strong liquids joins two critical points. Glass phase above $T_g$ becomes "ordered" liquid phase disappearing at $T_{LL}$ at low pressure and at $T_{n+}$=1.302×$T_m$ at high pressure.

## 1- Introduction:

Multiple liquid-to-liquid phase transitions (LLPTs) that are observed in several metallic glass-forming melts, have already been predicted using a classical nucleation equation completed by an enthalpy difference of two liquid phases depending on the square of the reduced temperature $\theta$ =(T-$T_m$)/$T_m$, where $T_m$ is the melting temperature [1]. The objectives of this paper are to extend the application of this renewed equation to the thermodynamic properties of water, to explain the occurrence of glass-to-glass phase transitions in amorphous water, and to show that the low-density phases obtained under decompression are glass phases analogous to those produced by vapour deposition at temperatures close to $T_g$ [2-5].

First-order transformations under pressure, induce high density amorphous phase [6-14], because the pressure increases the enthalpy and facilitates the glass transformation towards an equilibrium phase of higher density. These enthalpy and entropy changes cannot exceed in principle the value of the frozen enthalpy and entropy [15]. Any glass freezes enthalpy and entropy below $T_g$, which are available for exothermic relaxation or first-order transitions.

The water glass state is obtained by vapor deposition, liquid hyperquenching, confined water cooling, and application of high pressure to ice followed by various relaxation annealing [16]. A warming of bulk, amorphous water produces an endothermic event just below the crystallization temperature, occurring at around 136 K. [17-19]. The glass transition is characterized by a specific heat jump preceding the occurrence of crystallization. Applying high pressure to ice, reduces $T_m$ and produces amorphous water. Sharp transformations, viewed as first-order transitions are observed under pressure at 77 K, as well as at higher temperatures. These findings show that a bulk, amorphous liquid that has a low density can be transformed under pressure into a high-density amorphous liquid [6,9,18]. Three amorphous states have been identified under pressure: low density amorphous (LDA); high density amorphous l (HDA); and very high density amorphous (VHDA) after quenching to 77 K. Transformations of LDA to HDA and VHDA are observed after pressure is increased up to 16 kbar, and decompression taken down to residual pressures at temperatures lower than $T_g$. Some VHDA, HDA and LDA have been studied after complete decompression, down to temperatures of 77 K [7,10,20]. HDA obtained after decompression down to 100 bar and at a temperature of 77 K is also transformed by heating at ~140 K in LDA [7]. HDA is recovered by a new compression at p=0.32 GPa.

Measurements of water confined within silica gel in 1.1nm pores show the existence of a broad and high specific heat peak, at 227.5 K. (-45.6°C), and two heat flow changes at 124–136 K and 163–173 K, indicating the presence of two glass transitions [21,22]. A pronounced minimum of compressibility is still observed in water at a temperature of +45.5 °C, which is symmetrical with regard to $T_m$ at the ambient pressure of the transition at -45.6 °C [23,24]. A sharp specific heat increase below 273 K (already equal to 30 J.K.$^{-1}$mole$^{-1}$ at 235 K) is still observed in bulk supercooled water at ambient pressure, down to the crystallization temperature [16,25]. This confirms the possible existence, in the absence of crystallization, of a LLTP at temperatures smaller than 235 K [21,26].

In the first model (of two liquids), these phenomena are attributed, to the existence of a critical point leading to a line of first-order LLPTs [23,27-31]. The two liquids have the same chemical composition and contain low- and high-density species forming differently bonded domains. Such LLPTs form part of the general phenomenology for a wide range of liquids [32]. These ideas have successfully explained the existence of LLPTs, but are not able to predict glass thermodynamic properties, because they view them as a result of freezing, instead of a thermodynamic transition related to the difference of enthalpy of Phase 1 and Phase 2. The LLPT at 227.5 K looks like a first-order transformation of strong glass to fragile liquid [28].

In this paper, the glass transition is viewed as having a thermodynamic origin. There are many models describing it as a true phase transformation and experimental evidence favors this interpretation. The glass transition is seen as a manifestation of critical slowing down near a second-order phase transition with the possible existence of several classes of universality [33]. A model predicting the specific heat jump is based on a percolation-type phase transition with formation of dynamical fractal structures near the percolation threshold [34-40]. Macroscopic percolating clusters formed at the glass transition have been visualized [40]. High precision measurements of third- and fifth-order, non linear dielectric susceptibilities lead to a fractal dimension $d_F$=3 for the growing transient domains [41]. An observation of structural characteristics of medium-range order with neutrons and X-rays, leads to $d_F$=2.31 [42]. Another model, entirely based on thermodynamics, predicts the specific heat jump of strong and fragile glasses and liquid-to-liquid phase transitions [1,43]. For that, the classical nucleation equation is completed by introducing the enthalpy savings $-\varepsilon_{ls} \times \Delta H_m$, $-\varepsilon_{gs} \times \Delta H_m$, and $-\Delta \varepsilon_{lg} \times \Delta H_m$, respectively, associated with the growth critical nucleus formation leading to Phase 1 and Phase 2 above $T_g$ and Phase 3 below $T_g$, where $\Delta H_m$ is the melting heat [43]. The enthalpy difference $\Delta\varepsilon_{lg} \times \Delta H_m$, associated with the formation of vitreous Phase 3 below $T_g$, then equal to $(\varepsilon_{ls}-\varepsilon_{gs}) \times \Delta H_m$. The coefficients $\varepsilon_{ls}$ and $\varepsilon_{gs}$ are linear functions of $\theta^2 = (T-T_m)^2/T_m^2$, as shown by studying supercooling rate maxima of liquid elements [44,45]. A positive sign of $\Delta\varepsilon_{lg} = (\varepsilon_{ls}-\varepsilon_{gs})$ above $T_g$ and $T_m$ at a reduced temperature $\theta$ shows that Phase 1 is favored; a negative value would indicate that it is Phase 2 [1]. The first-order transition to a glass of confined liquid helium under pressure has been described using $\varepsilon_{ls0}=\varepsilon_{gs0}=0.217$ [45]. This glass is ultrastable if there is no more enthalpy to relax in this state. These values of $\varepsilon_{ls0}$ and $\varepsilon_{gs0}$, determined in many pure liquid elements at their melting temperature ($T_m$), correspond to the Lindemann coefficient 0.103 [46]. The transformation temperature ($T_{sg}$) of fragile glasses in ultrastable phases with higher density has been defined as a function of an enthalpy excess $\Delta\varepsilon \times \Delta H_m$ frozen after quenching. The denser ultrastable glass attains its lowest enthalpy at a transformation temperature $T_{sg}$ for a value of $\Delta\varepsilon$ equal to the frozen enthalpy of the glass below $T_g$ [5]. Phase 3 can be transformed into polyamorphous phases producing sharp enthalpy changes at various temperatures ($T_{sg}$) depending on smaller $\Delta\varepsilon$. The most-ultrastable phase is up to now the fully-relaxed glass.

## 2- Basic equations applied to water

The completed nucleation equation is given by (1):

$$\Delta G = \frac{4\pi R^3}{3}\Delta H_m / V_m \times (\theta - \varepsilon) + 4\pi R^2 (1+\varepsilon)\sigma_1 \qquad (1)$$

where $\Delta G$ is the Gibbs free energy change per volume unit, (associated with the formation of a spherical growth nucleus of radius R), $\varepsilon$ is a fraction of the melting enthalpy $\Delta H_m$ (equal to $\varepsilon_{ls}$ for a nucleus of Phase 1, $\varepsilon_{gs}$ for a nucleus of Phase 2, $\Delta\varepsilon_{lg}$ for a nucleus of Phase 3), $V_m$ is the molar volume, and $\theta = (T-T_m)/T_m$ is the reduced temperature. The melting heat $\Delta H_m$ and $T_m$ are assumed to be the same, whatever the nucleus radius R is, and not dependent on R. The critical nucleus can give rise to Phase 1, Phase 2, glass Phase 3, or various LLPT, according to the thermal variations of $\varepsilon$. The new surface energy is $(1+\varepsilon) \times \sigma_1$ instead of $\sigma_1$. The classical equation is obtained for $\varepsilon=0$ [47]. The homogeneous nucleation temperatures are $\theta_{n-}=(\varepsilon-2)/3$ for $\theta<0$ and $\theta_{n+}=\varepsilon$ for $\theta>0$ [1,43]. The critical radius is infinite at the homogeneous nucleation temperature obtained for $\theta=\varepsilon$ instead of $\theta=0$ for the classical equation. A catastrophe of nucleation occurs at $\theta=\varepsilon$ for crystals protected against surface melting [48].

The coefficients $\varepsilon_{ls}$ and $\varepsilon_{gs}$ in equations (2) and (3) represent values of $\varepsilon(\theta)$, and lead to the nucleus formation having the critical radius for Phase 1 and Phase 2 supercluster formations under pressure:

$$\varepsilon_{ls}(\theta) = \varepsilon_{ls0}(1-\theta^2 \times \theta_{0m}^{-2}) + P_1, \qquad (2)$$

$$\varepsilon_{gs}(\theta) = \varepsilon_{gs0}(1-\theta^2 \times \theta_{0g}^{-2}) - \Delta\varepsilon + P_2, \qquad (3)$$

where $\Delta\varepsilon$ is the coefficient of enthalpy excess in Phase 2 being frozen after quenching Phase 1; $P_1=(p-p_0) \times V_{m1}/\Delta H_m$ and $P_2=(p-p_0) \times V_{m2}/\Delta H_m$ are the contributions of the pressure (p) to the enthalpy coefficients $\varepsilon_{ls}$ and $\varepsilon_{gs}$, and $p_0$ is the ambient pressure [5]. The coefficients $\varepsilon_{ls}$ and $\varepsilon_{gs}$ are equal to zero at the reduced temperatures $\theta_{0m}$ and $\theta_{0g}$ for $\Delta\varepsilon=0$ and $P_1=0$, and they correspond to the Vogel-Fulcher-Tammann temperatures above and below $T_g$, respectively. Equations (2) and (3) are applicable at the homogeneous nucleation temperatures $\theta_{n-}$ in Phase 1 and Phase 2 respectively. Equation (4) determines $\theta_{n-}$ for Phase 2, combining (3) with $\theta_{n-}=(\varepsilon_{gs}-2)/3$ [43]:

$$\theta_{n-}^2 \times \varepsilon_{gs0} \times \theta_{0g}^{-2} + 3\theta_{n-} + 2 - \varepsilon_{gs0} + \Delta\varepsilon - P_2 = 0 \tag{4}$$

The solutions for $\theta_{n-}$ are given by (5):

$$\theta_{n-} = (-3 \pm \left[9 - 4(2 - \varepsilon_{gs0} + \Delta\varepsilon - P_2)\varepsilon_{gs0} / \theta_{0g}^2\right]^{1/2})\theta_{0g}^2 / (2\varepsilon_{gs0}) \tag{5}$$

$\theta_{n-}$ of Phase 2 for the sign + is called $\theta_2$, given by (5).

Equation (6) determines the homogeneous nucleation temperature $\theta_{n-}$, for Phase 1, combining (2) at this temperature with $\theta_{n-}=(\varepsilon_{ls}-2)/3$:

$$\theta_{n-}^2 \times \varepsilon_{ls0} \times \theta_{0m}^{-2} + 3\theta_{n-} + 2 - \varepsilon_{ls0} - P_1 = 0 \tag{6}$$

The reduced homogeneous nucleation temperature $\theta_{n-}$ of Phase 1 under pressure in (7) is deduced from (6):

$$\theta_{n-} = (-3 \pm \left[9 - 4(2 - \varepsilon_{ls0} - P_1)\varepsilon_{ls0} / \theta_{0m}^2\right]^{1/2})\theta_{0m}^2 / (2\varepsilon_{ls0}) \tag{7}$$

$\theta_{n-}$ in (7) is called $\theta_1$ for the sign +. The glass transition occurs at $\theta_g$ when $\varepsilon_{ls}(\theta)$ in (2) is equal to $\varepsilon_{gs}(\theta)$ in (3). $\theta_1$ and $\theta_2$ are equal to $\theta_g$ in strong glasses because $\varepsilon_{ls}(\theta_g)=\varepsilon_{gs}(\theta_g)$ for $\Delta\varepsilon=0$ and $P_1=P_2=0$.

As water is a strong glass at low temperatures, the coefficients $\varepsilon_{gs0}$ in equation (8) and $\varepsilon_{ls0}$ in equation (9), deduced from equation (4) with $P_2=0$ and from equation (6) with $P_1=0$ and $\Delta\varepsilon=0$, are determined from the knowledge of $\theta_g$, $\theta_{0g}$, and $\theta_{0m}$ [43]:

$$\varepsilon_{gs0} = \frac{3\theta_g + 2}{1 - \theta_g^2 / \theta_{0g}^2} \tag{8}$$

$$\varepsilon_{ls0} = \frac{3\theta_g + 2}{1 - \theta_g^2 / \theta_{0m}^2} \tag{9}$$

where the reduced temperatures $\theta_{0g}$ and $\theta_{0m}$ are equal to -1 and -2/3, respectively, because the Vogel-Fulcher Tamman temperatures are equal to 0 K below $T_g$ and to $T_m/3$ (above $T_g$) for many pure, strong liquid elements [44]. With $T_g$=136.6 K, $\theta_g$=-0.5, $\varepsilon_{gs0}$ is equal to 0.66667 and $\varepsilon_{ls0}$ to 1.14286. The frozen enthalpy at $T_g$ is equal to the minimum value -0.3704×$\Delta H_m$ of ($\varepsilon_{ls}-\varepsilon_{gs}$)×$\Delta H_m$ obtained for $\varepsilon_{ls}$=0 at $\theta=\theta_{0m}$= -2/3 without imposing any entropy constraint [15]. The heat capacity jump at $T_g$ is equal to: $(d\varepsilon_{ls}/dT-d\varepsilon_{gs}/dT)\times\Delta H_m$=-1.905*$\Delta H_m/T_m$=41.9 JK$^{-1}$mole$^{-1}$ in agreement with old measurements [18], as shown in Figure 1. The specific heat excess $\Delta C_p(T)=d(\varepsilon_{ls}-\varepsilon_{gs})/dT\times\Delta H_m$ of supercooled liquid only exists above the Kauzmann temperature because the entropy excess of supercooled liquid cannot be larger than the fusion entropy $\Delta S_m$. $\Delta C_p(T)$ is used to evaluate the Kauzmann temperature ($T_K$) of supercooled water. The entropy excess ($\Delta S_m$) of supercooled water is equal to $\Delta H_m/T_m=6000/273.1=22$ J.$K^{-1}$mole$^{-1}$ between 227.5 K and T=116.5 K and between 119.7 K and 273.1K. The Kauzmann temperature occurs at $T_K\cong116.5$ K-119.7 K.

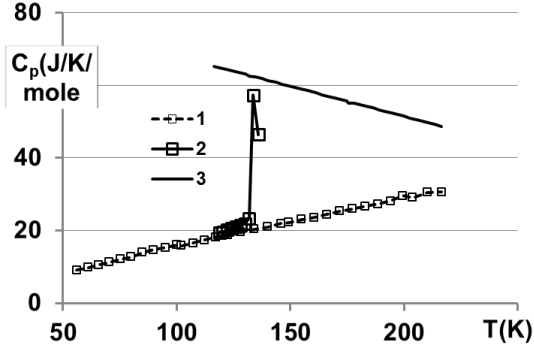

**Figure 1**: *1. Heat capacity of hexagonal ice [18] 2. Heat capacity jump at 133.6 K [18]; 3. Supercooled water heat capacity calculated with the derivative $(d\varepsilon_{ls}/dT - d\varepsilon_{gs}/dT)\times\Delta H_m$.*

For confined supercooled water in pores of radius R=0.55 nm [21,22], $\theta_2$ is equal to -0.167 (227.5 K) for $P_2$=0.8505 and $\Delta\varepsilon$=0 in equation (5). Using the Young-Laplace equation, $\Delta p$ is equal to $2\gamma/R=0.31\pm0.02$ GPa, with a value of the surface tension $\gamma=0.085\pm0.005$J/m$^2$ at 227.5 K, extrapolated from its thermal variation above 250 K [23]. The enthalpy coefficient ($P_2$) is deduced to be close to 0.8505 with $V_{m2}\cong16.5\times10^{-6}$ m$^3$ and p=0.31 GPa.

The enthalpy difference coefficient $\Delta\varepsilon_{lg}$ between vitreous Phase 3 and liquid Phase 1 is given by equation (10) under pressure (p):

$$\Delta\varepsilon_{lg}(\theta) = (\varepsilon_{ls} - \varepsilon_{gs}) = \varepsilon_{ls0} - \varepsilon_{gs0} + \Delta\varepsilon + P_1 - P_2 - \theta^2(\varepsilon_{ls0}/\theta_{0m}^2 - \varepsilon_{gs0}/\theta_{0g}^2) \qquad (10)$$

The difference $\Delta P=(P_1-P_2)=\delta V\times p/\Delta H_m$ is proportional to the volume change ($\delta V$), and to the pressure (p) at the transformation temperature ($\theta$). ($\Delta P$) is equal to zero for $\delta V$=0 in the absence of latent heat. The homogeneous nucleation temperature of Phase 3 also occurs for $\Delta\varepsilon_{lg}$=0 with $\Delta\varepsilon$=0 because Phase 1 and Phase 2 have the same homogeneous nucleation temperature $\theta_1=\theta_2=\theta_g$.

A sharp enthalpy difference between non-relaxed glass Phase 3 and fully-relaxed glass Phase 3 can be induced in all glasses below $T_g$ for $\Delta\varepsilon_{lg}$=0 in (10), when an enthalpy excess coefficient ($\Delta\varepsilon$) exists after rapid cooling, as already described for an ultrastable glass formation [5,10]. This enthalpy difference is equal to $-2\times\Delta\varepsilon_{lg}(\theta)\times\Delta H_m$ above $\theta_K$ for $\Delta\varepsilon$=0 and $P_1=P_2$=0 because it cannot exceed the frozen enthalpy which is available at any temperature below $T_g$. This transformation temperature $T_{sg}$ for a stable glass formation given in (11) is equal to, or larger than, $T_K$ and is also induced by pressure. It depends on $\delta V$ and on the value of $\Delta\varepsilon$ at this temperature.

$$\theta_{sg} = -\left[\left(\varepsilon_{ls0} - \varepsilon_{gs0} + \Delta\varepsilon + \Delta P\right)/\left(\varepsilon_{ls0}\theta_{0m}^{-2} - \varepsilon_{gs0}\theta_{0g}^{-2}\right)\right]^{1/2} \tag{11}$$

After decompression, the enthalpy change coefficients of supercooled water are represented in Figure 2. The line $\Delta\varepsilon_{lg}$=0 at the origin corresponds to a quenched glass phase, containing a positive enthalpy excess which is equal to $\Delta\varepsilon\times\Delta H_m$=[-1.14286×(1-2.25×$\theta^2$)+0.66667×(1-$\theta^2$)]×$\Delta H_m$ above the line 1 because $\Delta\varepsilon_{lg}$ of the glass phase is negative for $\Delta\varepsilon$=0. The total enthalpy of this quenched phase is then equal to that of Phase 1. The nonrelaxed glass phase is represented by Line 1 and the fully-relaxed phase at thermodynamic equilibrium by Line 2. A sharp, spontaneous transition is observed at 117 K during heating of the HDA phase [10] and this corresponds to an enthalpy coefficient excess $\Delta\varepsilon$=0.146 at $\theta$=-0.5735 as shown in Figure 2. The latent heat measured using continuous heating is 757±144 J.mole⁻¹ [6] and it corresponds to the relaxation of an enthalpy excess equal to 0.146×$\Delta H_m$=876 J.mole⁻¹. An isothermal relaxation at $T_{sg}$=117 K would have to deliver a latent heat two times larger (and equal to 1752 J.mole⁻¹) as shown in Figure 2. The sample volumes of HDA in Figure 3 have been also measured at p=0 after a duration of about 3 hours of isotherm annealing [10]. In these conditions, a sharp volume change of 0.16×10⁻⁶ m³g⁻¹ occurs at 115±0.5 K confirming the existence of spontaneous and high enthalpy relaxation of about 1752 J.mole⁻¹ from HDA to LDA phases. Following this analysis, this transition temperature at $T_{sg}$ is the first observation of a glass Kauzmann temperature ($T_K$) because spontaneous and sharp enthalpy relaxation is only possible above $T_K$. Latent heats are still produced at various temperatures ($T_{sg}$) depending on $\Delta\varepsilon$ and they correspond to partial relaxations of Phase 3. This type of transition has already been observed in other glasses obtained by vapour deposition on substrates maintained at the temperature $T_{sg}$ using very slow deposition rates [2,4] and described by the same model [5]. This analysis is based on the existence of two main glass phases (Phase 3), the first one being the nonrelaxed classical glass phase with its frozen enthalpy equal to -0.37037×$\Delta H_m$ and the second one the ultrastable glass phase which is expected to be fully relaxed with a maximum enthalpy reduction equal to -0.2923×$\Delta H_m$ (at approximately $T_K$≅117 K).

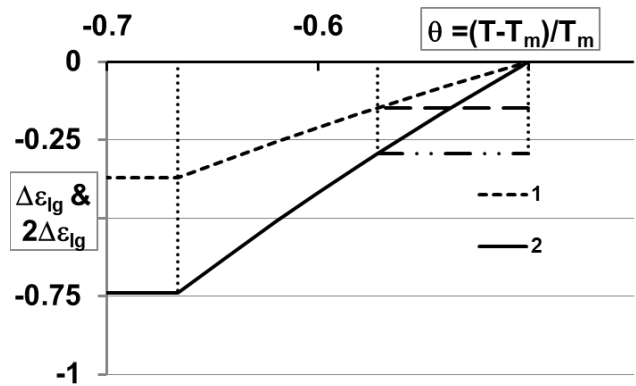

**Figure 2**: *Prediction of reduced temperatures $\theta_{sg}$ along the line $\Delta\varepsilon_{lg}$=0 and latent heat coefficients $\Delta\varepsilon_{lg}$ associated with the glass-to-glass transformations at zero pressure. Lines numbered from 1 to 2: 1. $\Delta\varepsilon_{lg}(\theta)$ given by equation (10) of nonrelaxed Phase 3 with $\Delta\varepsilon$=0, $\Delta P$=0, $\theta_g$=-0.5 ($T_g$=136.6 K); **2**. Equilibrium enthalpy coefficient $2\times\Delta\varepsilon_{lg}(\theta)$ of fully-relaxed Phase 3 crossing $\theta_g$=-0.5 ($T_g$=136.6 K). Observed reduced*

*temperature $\theta_{sg}$=-0.5735, corresponding to $T_{sg}$=$T_K$≅117 K, accompanied by latent heats equal to 877Jmole$^{-1}$ ($\Delta\varepsilon_{lg}$=-0.14615) or 1752 J.mole$^{-1}$ (2×$\Delta\varepsilon_{lg}$=-2×0.14615=-0.2923).*

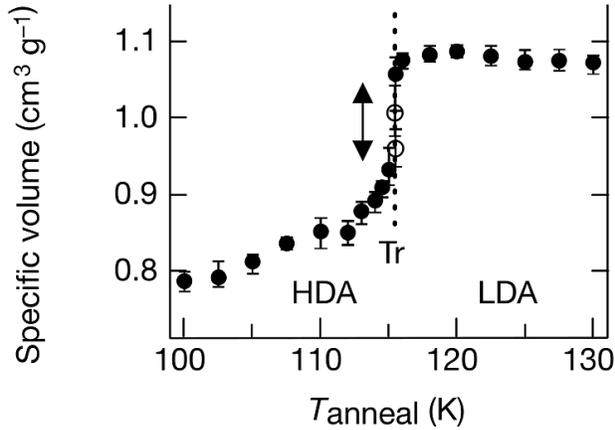

**Figure 3**: *Reproduced from [10] with Nature Publishing Group permission. Isotherm annealing of HDA at different temperatures over 3 hours' duration. Specific volume versus annealing temperature.*

Sharp, exothermic latent heats are still observed in water after decompression of VHDA at 77K from various pressures, see Figure 4 [20]. VHDA under pressure has a much larger density than HDA. The glass transition at 136.6 K and after decompression is not detected in these samples (Figure 4). The sharp, exothermic latent heats observed below 136 K decrease the density and give rise to ice which contains orientational disorder instead of fully relaxed Phase 3 [49]. The latent heat at the crystallization temperature of 164 K seems to depend on the preceding exothermic heat. The absence of glass transition at 136.6 K confirms that the sharp transitions of polyamorphous phases at p=0 [20] lead to amorphous ice resulting from molecular reorientation processes [49]. The enthalpy coefficient along Line 2 in Figure 2 cannot be attained without amorphous ice formation. Sharp exothermic latent heats are observed around $T_{sg}$=125, 126, 130, 132, 134 K and predicted to be equal along Line 2 to 1000, 876, 562, 383, 217 J.mole$^{-1}$ respectively. The latent heats at 130, 132, and 134 K are in rough agreement with those observed in Figure 4, whereas those at 125 and 126.5 K are smaller because the annealing time at these temperatures is too small during a continuous heating.

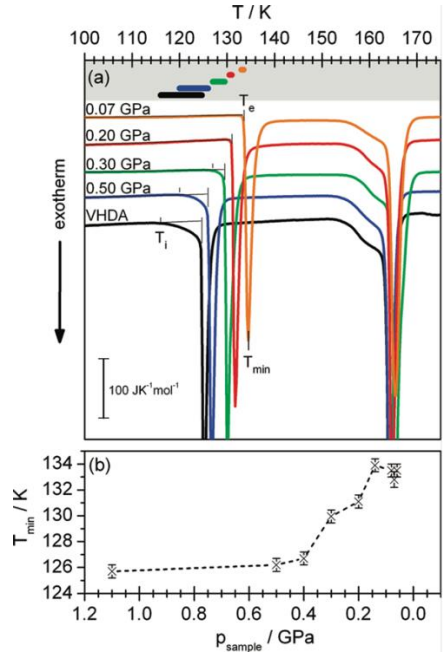

**Figure 4**: *Reprinted from [20, Figure 3] with ACS permission . " (a) DSC scans recorded at a rate of 10 K/min. The DSC output signal was normalized to 1 mol. The samples were heated from 93 to 253 K; thermograms are plotted in the temperature range of 100–175 K. Shown are VHDA (black line) and four samples made by decompression of VHDA at 140 K to 0.5 (blue line), 0.3 (green line), 0.2 (red line), and 0.07 GPa (orange line). First exothermic peak: transition to LDA; second exothermic peak: crystallization to cubic ice. The bars in the top part indicate the difference between $T_i$ and $T_e$, which is a measure of the relaxation state of the sample.(b) The temperature at the peak minimum $T_{min}$ is shown as a function of the pressure from which the sample has been recovered. The values are not only obtained from the four selected runs depicted in (a) but represent mean values obtained from several runs" There is no detectable endothermal event at T=136.6 K.*

There are two regions of water crystallization (at approximately 230–250 K and 135–165K) confirming the existence of a LLPT between this two regions in all the samples studied. Supercooled water undergoes a first-order phase transition that separates fragile from strong states. Fragile liquids have values of $\varepsilon_{ls0}$ given in (12) [43]:

$$\varepsilon_{ls}(\theta = 0) = \varepsilon_{ls0} + P_1 = 1.5 \times \theta_1 + 2 + P_1 = a \times \theta_g + 2 + P_1, \tag{12}$$

where a=1 leads to a specific heat excess $\Delta C_p(T)$ of the supercooled melt at the glass transition equal to $1.5 \times \Delta H_m/T_m$ [50-52]. The reduced temperature $\theta_{0m}$ is given by (13), and is a double solution for (6):

$$\theta_{0m}^2 = \frac{8}{9}\varepsilon_{ls0} - \frac{4}{9}\varepsilon_{ls0}^2. \tag{13}$$

New parameters $\varepsilon_{gs0}$ and $\theta_{0g}$ are fixed at $T_g$ and below $T_g$ in equations (14) and (15) to give a double solution for (4) with a=1 because a<1 leads to a too high nucleation temperature ($T_1$) in Phase 1; $\varepsilon_{gs0}$ is maximized by (14) and (15) [43]:

$$\varepsilon_{gs}(\theta = 0) = \varepsilon_{gs0} + P_2 = 1.5 \times \theta_g + 2 + P_2, \tag{14}$$

$$\theta_{0g}^2 = \frac{8}{9}\varepsilon_{gs0} - \frac{4}{9}\varepsilon_{gs0}^2. \qquad (15)$$

The first-order transition under Laplace pressure of fragile-to-strong water in confined space, occurs at $T_{LL}$=227.5 K, $\theta_{LL}$=-0.167, and $T_m$=273.1 K for the melting temperature of superclusters percolating in the glass state as assumed in equation (1) [22]. The two temperatures, where $\Delta\varepsilon_{lg}$ in equation (10) is equal to zero, cannot depend on the pressure because there is no volume change there. In the fragile state of water, these two reduced temperatures are $\theta_{LL}$=±0.16705 and they are symmetrical with regard to $T_m$ in Figure 5. For $\Delta\varepsilon_{lg}$=0 above $T_m$, there is a compressibility minimum of bulk water at 45.6°C because Phase 2 replaces Phase 1 without volume change at this temperature as shown in Figure 5 [23,53,54]. Then, the first-order transition of fragile-to-strong liquid does not depend on the pressure at $\theta_{LL}$=-0.167. The value a=1 used in Figure 5 leads to $\varepsilon_{ls0}$=1.7953, $\varepsilon_{gs0}$=1.69295, $\theta_{0g}^2$=0.23103, $\theta_{0m}^2$=0.16333 with $\theta_g$=-0.2047 and to $T_1$=235.9 K in good agreement with a maximum supercooling of water equal to 35 K. The coefficients $\varepsilon_{ls}$ and $\varepsilon_{gs}$ of fragile liquids are equal to 1.5 at $\theta=\theta_{LL}$ while $\varepsilon_{ls}$ and $\Delta\varepsilon_{lg}$ of strong liquids are respectively equal to 1.07102 and 0.42298. Under these conditions, the water specific heat increase up to 81 J.K.$^{-1}$ mole$^{-1}$ at $\theta_{LL}$=-0.167 [22], is due to an LLPT [23,27-30]. This increase is not only observed at zero pressure, but also under a Laplace pressure of 0.31±0.02 GPa in 1.1 nm pores, slightly increasing with decreasing temperature [22,55-57]. In Figure 5, the LLPT at 227.5 K and zero pressure is accompanied (during heating) by an exothermic latent heat of (-1.5+1.07102+0.42298)×$\Delta H_m$=-0.006×$\Delta H_m$ associated with Phase 1 and Phase 3 transformations because glass Phase 2 is transformed in liquid Phase 3 at $T_g$ and continues to exist as a liquid phase above $T_g$. As expected [22-28], a critical point seems to exist for p=0 in these conditions.

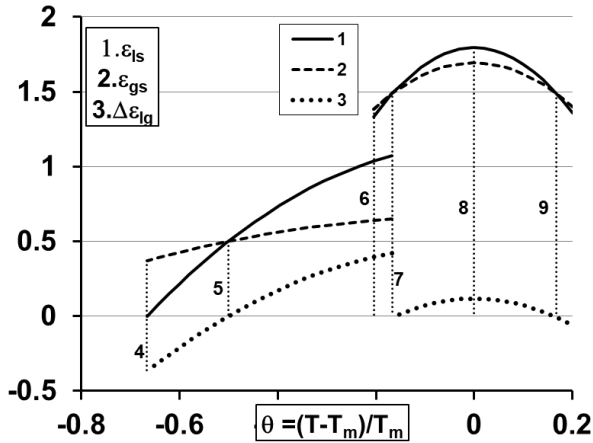

**Figure 5: The enthalpy coefficients $\varepsilon_{ls}$, $\varepsilon_{gs}$ and $\Delta\varepsilon_{lg}$ of Phase 1, Phase 2 and Phase 3, P=0.** *For the strong liquids at $\theta_{LL}$=-0.167, $\varepsilon_{ls}$ =1.07102, $\varepsilon_{gs}$=0.64804, $\Delta\varepsilon_{lg}$=0.42298 and at $\theta_g$=-0.5, $\varepsilon_{ls}=\varepsilon_{gs}$=0.5. For the fragile liquids at $\theta_{LL}$=-0.167, $\varepsilon_{ls} = \varepsilon_{gs}$=1.5. Curves numbered from 1 to 3: 1- $\varepsilon_{ls}(\theta)$ of Phase 1 given by (2) below $\theta_{LL}$=-0.167 in the strong water and above in the fragile one; 2- $\varepsilon_{gs}(\theta)$ of Phase 2 given by (3) below and above $\theta_{LL}$=-0.167; 3- $\Delta\varepsilon_{lg}(\theta)$ of Phase 3 below and above $\theta_{LL}$. Temperatures numbered from 4 to 9: 4- The frozen enthalpy coefficient $\Delta\varepsilon_{lg}$ equal to -0.3704 at $\theta$=-2/3; 5- The glass transition $\theta_g$=-0.5 ($T_g$=136.6 K); 6- $\theta_g$=-0.2047 of the fragile water in the absence of LLTP; 7- At $\theta_{LL}$=-0.167, $T_{LL}$=227.5 K,*

*the LLTP for Δε_lg=0; 8- θ=0 the melting temperature at T_m=273.14 K; 9- For θ= 0.167, Δε_lg=0 at the isothermal compressibility minimum temperature 318.7K.*

### 3- First-order transformations LDA-HDA under pressure

High pressure applied to ice samples followed by complete decompression, induces an LDA phase which has a density and, consequently, an enthalpy close to that of ice [6]. The LDA phase is viewed as having an enthalpy difference of $-0.3704 \times \Delta H_m$ with that of the glass phase at all temperatures $T<T_g$. The LDA to HDA involving high pressures and an enthalpy excess $(\Delta \varepsilon)$ due to Phase 1 quenching is frozen because the melting temperature $(T_m)$ is strongly decreased, and the sample is cooled during decompression from temperatures much higher than $T_m$. In Figure 6, compression experiments of this LDA phase at various temperatures, transform it into an HDA phase at a well-defined pressure (p). This sharp transformation is also viewed as an HDA to LDA transformation because this first-order transition is reversible at $T_{sg}$. The volume change $\delta V$ in (10) does not depend on the pressure (p) and is equal to $0.2 \times 10^{-6}$ $m^3g^{-1}$, as shown in Figure 6 [8].

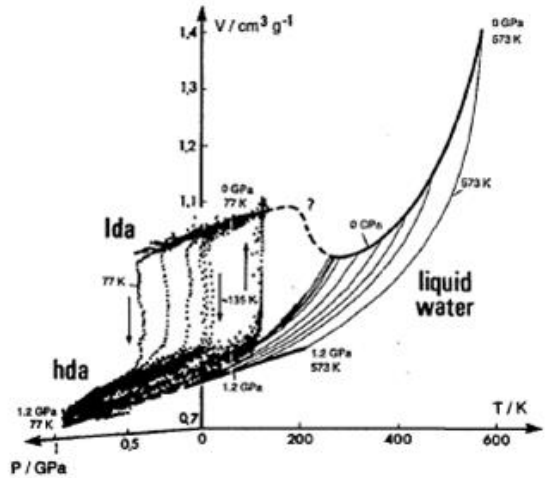

**Figure 6**: **Specific volume versus pressure (GPa) and Temperature (K)**. *Reprinted from [8, Figure 4] with AIP permission. "Low density amorphous to high density amorphous transformations under pressure occur for p=0.55, 0.45, 0.38, 0.32, 0.05 GPa and T=77, 100, 121, 135, 140 K respectively. Liquid water under 1.2 GPa and 0 pressures is also represented versus temperature. The linkage between LDA and liquid state at zero pressure occurs at 227 K. The liquid at zero pressure corresponds to Phase 1"*

All values of various quantities are given in Table 1. The sharp enthalpy changes under pressure (p) are equal to $\Delta P_2=0.3704 \times \Delta H_m$ and occur at $\theta=\theta_{sg}$, as given by equation (11). LDA is viewed as having the enthalpy of nonrelaxed glass Phase 3 and an effective enthalpy excess $\Delta \varepsilon_{eff}$, and it is expected to have an enthalpy difference $-\Delta P_2$ with HDA for $T_{sg}<T<T_g$. This LDA-HDA first-order transformation is subject to an enthalpy constraint setting that the total enthalpy increase at equilibrium cannot be larger than the maximum frozen enthalpy $0.3704 \times \Delta H_m$ (produced at $\theta=-2/3$). In Figure 7, the values of $\Delta \varepsilon_{eff}=\Delta \varepsilon+\Delta P_1$ given in Table 1 (obtained using (10) for $\theta=\theta_{sg}$ and represented as a function of $\theta^2_{sg}$), depend on pressure via $\Delta P_1$ which is the initial enthalpy change under pressure associated with the glass volume change below $T_g$, before the occurrence of the first-order transition. The values of $\Delta\varepsilon_{eff}$ at the temperature $\theta_{sg}$ are negative because $\Delta P_1$ is negative. The enthalpy excess $\Delta\varepsilon$, in the absence of pressure, is equal to $(-\varepsilon_{lg}+\varepsilon_{gs})$ and varies from 0.3704 at T=77 K to zero at 136.6 K, as shown in Table 1. Enthalpy excess depends on the reduced temperature $\theta$ below 136.6 K, as has already been observed in hyperquenched glasses below $T_g$ [58-60]. The values of $\Delta P_1$ are deduced from the difference: $\Delta\varepsilon_{eff}$-$\Delta\varepsilon$. The melting temperatures ($T_m$) under pressure are assumed to be those of hexagonal ice [9]. They lead in water to the maximum change of the enthalpy coefficient $\Delta\varepsilon_{lg}$=0.3704 at the transformation temperature $T_{sg}$ and for 0.31<p≤0.6 GPa. The first-order transformation of LDA-HDA, takes into account the entropy constraint which could not be respected for an enthalpy relaxation.

**Table 1:** *1- p the applied pressure. 2- $T_{sg}$ the temperature of the first-order glass-to-glass transition. 3- $T_m$ the melting temperature depending on pressure. 4- $\Delta\varepsilon$ the enthalpy excess coefficient induced by quenching at $T_{sg}$. 5- $\Delta P_1$=p×$\delta V_1$/$\Delta H_m$ the change of the enthalpy coefficient associated with the volume difference between liquid and glass just before the transformation from LDA to HDA under pressure (p); $\Delta H_m$ the fusion heat. 6- $\theta_{sg}$=($T_{sg}$-$T_m$)/$T_m$. 7- $\Delta P_2$=p×$\delta V_2$/$\Delta H_m$ the change of the enthalpy coefficient associated with the volume difference induced by the first-order transition under pressure. 8- $\Delta\varepsilon_{eff}$=$\Delta P_1$+$\Delta\varepsilon$ the effective enthalpy excess coefficient under pressure. 9- $\theta_g$=($T_g$-$T_m$)/$T_m$ the reduced glass transition temperature of HDA under pressure p. 10- f the HDA entropy fraction from $T_{sg}$ to $T_g$ under pressure (p). 11- $\Delta S$ the entropy excess under pressure of HDA from $T_g$ to $T_m$. 12- $\theta_{gS}$ the reduced glass transition temperature of HDA under pressure calculated from the entropy constraint. 13- $T_{gS}$ the glass transition temperature of HDA under pressure deduced from the entropy constraint. 14- $T_g$ the glass transition temperature calculated from equation (5) assuming the enthalpy coefficient difference of HDA with LDA is ($\varepsilon_{gs0}$+$\Delta P_2$)=0.66667+0.3704=1.037 instead of $\varepsilon_{gs0}$ for p=0.*

| | | | | | |
|---|---|---|---|---|---|
| 1 | **p (GPa)** | 0.6 | 0.5 | 0.42 | 0.35 |
| 2 | **$T_{sg}$ (K)** | 77 | 100 | 121 | 135 |
| 3 | **$T_m$(K)** | 154 | 186 | 215 | 229 |
| 4 | **$\Delta\varepsilon$** | 0.3704 | 0.2892 | 0.1148 | 0 |
| 5 | **$\Delta P_1$** | -0.3704 | -0.3582 | -0.2295 | -0.1654 |
| 6 | **$\theta_{sg}$** | -0.5 | -0.4624 | -0.4372 | -0.4035 |
| 7 | **$\Delta P_2$** | 0.3704 | 0.3704 | 0.3704 | 0.3704 |
| 8 | **$\Delta\varepsilon_{eff}$** | 0 | -0.067 | -0.111 | -0.154 |
| | **No man's land** | | | | |
| 9 | **$\theta_g$** | -0.3677 | -0.3677 | -0.3677 | -0.3677 |
| 10 | **f=0.3704×$T_m$/$T_{sg}$** | 0.741 | 0.689 | 0.658 | 0.621 |
| 11 | **$\Delta S$(JK$^{-1}$mol$^{-1}$)** | 5.7 | 6.84 | 7.52 | 8.34 |
| 12 | **$\theta_{gS}$** | -0.325 | -0.352 | -0.366 | -0.383 |
| 13 | **$T_{gS}$ (K)** | 104 | 121 | 136 | 141.3 |
| 14 | **$T_g$ (K)** | 97.4 | 117.6 | 136 | 144.7 |

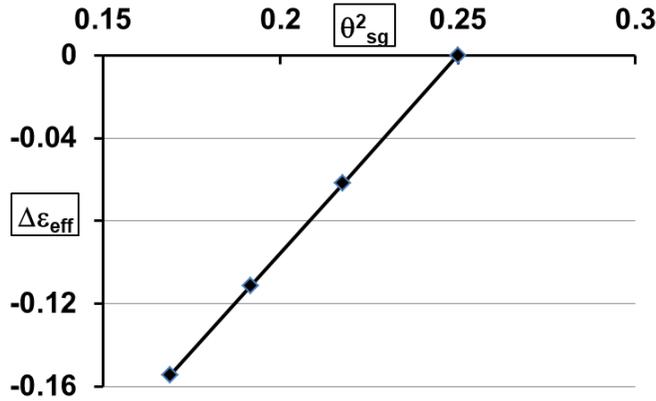

**Figure 7:** $\Delta\varepsilon_{\text{eff}}$ versus $\theta^2_{\text{sg}}$. *The values of $\Delta\varepsilon_{\text{eff}}$ and $\theta_{\text{sg}}$ are reported in Table 1.*

In Figure 8, the transformation under pressure starts from Line 3 and leads to HDA, including $\Delta\varepsilon_{\text{eff}}$ on line 1 which is characterized by an enthalpy increase equal to the frozen enthalpy $(0.3704\times\Delta H_m)$ below $T_g=136.6$K. The volume change ($\delta V=0.3704*\Delta H_m/18/p=0.206\times10^{-6}\text{m}^3\text{g}^{-1}\text{GPa}^{-1}$ for $p=0.6$GPa) is constant under various pressures (p), and equal to the experimental value presented in Figure 6. The enthalpy difference coefficient ($\Delta P_2=-0.3704$) is the sum of $\Delta\varepsilon_{\text{eff}}$ and $\Delta P_1$. $\Delta P_1$ is negative and proportional to the applied pressure in agreement with Figure 6. The slope $d\Delta\varepsilon_{\text{eff}}/dp$ corresponds to the measured value $(dV/dp=0.21\text{g}^{-1}\text{cm}^3\text{GPa}^{-1})$ of the LDA phase [13, Figure 14].

The LDA-to-HDA transformation, occurring for $p=0.35$ GPa in the interval 130–140 K, is reversible when the pressure is decreased down to $p=0.05$ GPa [8]. This reversibility still proves its first-order character. The enthalpy change induced at $T_{sg}$ is recovered near $T_g\cong0.5\times T_m$ after decompression down to 0.05 GPa. The enthalpy excess ($\Delta\varepsilon_{\text{eff}}\times\Delta H_m$) is fully recovered at $T=T_{sg}$ and $p=0$ because the pressure changes the enthalpy from Line 1 to Line 3, in Figure 8, and decreases the volume by a constant quantity which is recovered after decompression.

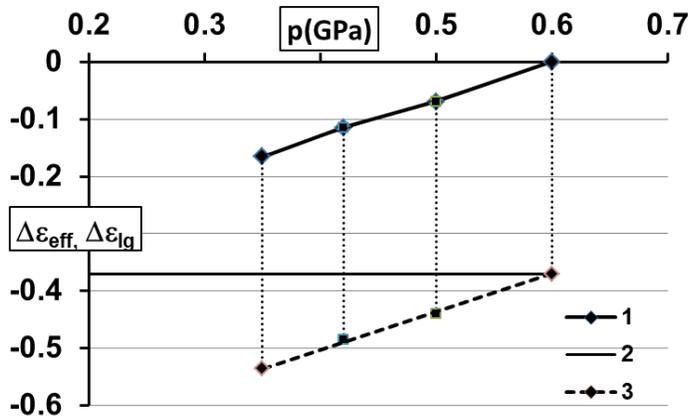

**Figure 8**: **Enthalpy excess coefficients Δε$_{eff}$ versus pressure p and sharp enthalpy coefficient changes Δε$_{lg}$ at θ=θ$_{sg}$:** 1– *The line 1 is the HDA line represented by Δε$_{eff}$ versus p (GPa). 2- The line 2 represents the change Δε$_{lg}$=-0.3704=ΔP$_2$ which is equal to Δε$_{eff}$+ΔP$_1$ given in Table 1. ΔP$_1$ is proportional to p. 3- The line 3 is the LDA line with a slope corresponding to dV/dp= 0.206 g$^{-1}$cm$^3$GPa$^{-1}$. Calculated points roughly correspond to Mishima's measurements reproduced in Figure 6 at T=77K for p=0.6 GPa instead of 0.55 GPa, for p=0.5 instead of 0.45 GPa, for p=0.42 instead of 0.38 GPa , and for p= 0.35 GPa instead of 0.32 GPa [8].*

A relaxation of the enthalpy excess Δε has been observed [22] for water confined into 1.1nm pores being submitted to Laplace pressure and this is reproduced in Figure 9. "The systematic heat-evolution and heat-absorption effects for the rapidly and slowly cooled samples are characteristic of a glass transition, and two transitions are found" between the ranges 124-136 K and 163-172 K. The glass transition of all LDA phases above T$_{sg}$ is given by (5) where Δε-P$_2$=0 and ε$_{gs0}$ is replaced by ε$_{gs0}$+ΔP$_2$=0.66667+0.37034=1.037 because ΔP$_2$ results from the first-order transition at T$_{sg}$ instead of relaxation. It is equal to θ$_g$=-0.3677 and T$_g$=0.632×T$_m$. With T$_m$=273.1 K, the T$_g$ of confined water is expected to occur at 173 K. Δε=-Δε$_{lg}$ disappears at θ$_g$=-0.5 (T$_g$=136.6 K) when Δε$_{lg}$=0. The relation Δε$_{eff}$=Δε+ΔP$_1$ shows that Δε$_{eff}$=ΔP$_1$ occurs at T≅136.6 K. This equality occurs in Figure 9 for ΔP$_1$=-0.3704/2=-0.185. The Laplace pressure at T≅136 K is then equal to 0.30 GPa in good agreement with 0.31 GPa at T$_{LL}$=227.5K.

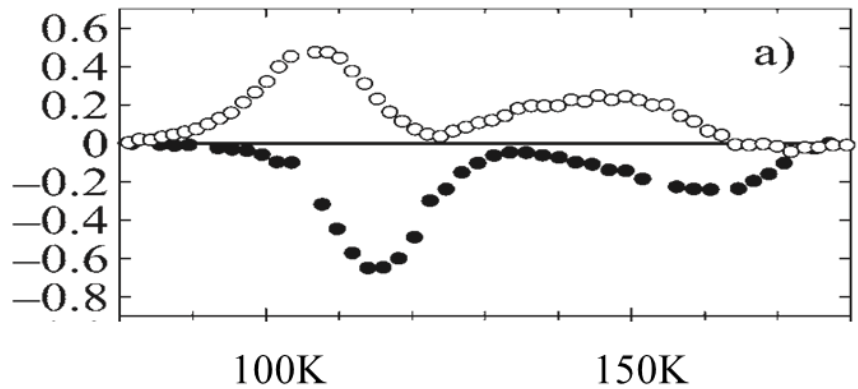

**Figure 9**: *Reproduced from [22] with AIP permission. "Temperature dependence of the rates of spontaneous heat release and absorption, observed in the heat-capacity measurements of ordinary water (H$_2$O) by an intermittent heating method. Average pore diameter: a) 1.1 nm, ○=sample cooled rapidly at around 5 K.min$^{-1}$ before the measurements, ●=sample cooled slowly at 10 mKmin$^{-1}$".*

The glass transition calculated by equation (5) occurs where θ$_g$=-0.5 (T$_g$=0.5×T$_m$) at low pressures in the absence of an LDA-HDA transition. The HDA phase has a larger enthalpy, and the new glass transition temperature of HDA is still equal to (5) and T$_g$=0.632×T$_m$ depending on the melting temperature (T$_m$) under pressure. For p=0.6 MPa, T$_m$=154 K, T$_g$≅97.4 K; for confined water in 1.1 nm pores, T$_g$=173 K which is in agreement with Figure 9.

The first-order LDA-HDA transition at $T_{sg}$ under pressure is accompanied by an entropy change equal to $0.3704 \times \Delta H_m / T_{sg} = f \times \Delta H_m / T_m$ where f is a fraction of the fusion entropy at the melting temperature under pressure which is recovered at the glass transition, $T_{gs}$:

$$f = 0.3704 \times T_m / T_{sg} \tag{16}$$

Values for f are given in Table 1. The entropy ($\Delta S$) of HDA at $T_{gS}$ in Table 1 is counted from $T_{gS}$ to the melting temperature $T_m$ using the specific heat, deduced from $d\Delta\varepsilon_{lg}/dT$, because the LLPT has disappeared above p=0.31 GPa (see section 3). Some glass transitions (under a pressure at $\theta_{gS}$ and $T_{gS}$ calculated using the entropy constraint) are given in Table 1 and they roughly equate to the calculated values $\theta_g$ and $T_g$ using equation (5). The uncertainty on $\Delta S$ values is estimated to be 8.6% from the difference between $\Delta S = 7.59$ J.K.$^{-1}$mole$^{-1}$ at $\theta_g$ and $\Delta S = 5.7$ J.K.$^{-1}$mole$^{-1}$ for p=0.6 GPa.

The first-order transition of HDA to LDA induces an enthalpy decrease equal to $-0.37037 \times \Delta H_m$ and then a very stable glass state up to $T_g = 136.6$ K. In addition, a spontaneous enthalpy relaxation of this state is observed at, or above, $T_K$ in order to attain the enthalpy of the ultrastable state after isothermal and complete relaxation of the frozen enthalpy at $T_{sg} = T_K$. This complete relaxation gives rise to orientational disorder in ice instead of ultrastable glass phase [49].

Recent studies have reported on in-situ structural characterization of LDA after decompression and relaxation between 96K and 160K by synchrotron x-ray diffraction [61]. An intermediate crystalline phase at 100K, prior to complete amorphization at 133K is observed. These results show that LDA exists under various forms depending on the relaxation temperature because any phase having a fusion entropy smaller than $\Delta H_m / T_m = 6000/273.1 = 22$ J.K.$^{-1}$mole$^{-1}$ can be condensed at a temperature equal or larger that its own Kauzmann temperature. Another publication classifies HDA as a "derailed" state along the ice Ih to high-density ice IV pathway [62]. These two papers show that the same volume changes that characterise LDA and HDA lead to various phases including "derailed" states depending on relaxation time and temperature before attaining ice. Nevertheless, relaxed LDA has an enthalpy still smaller than that already frozen below $T_g$. Its enthalpy is so close to that of ice that its vitreous state can be transformed, by relaxation, through various "derailed" states on the pathway leading to the formation of ice [49].

## 4- The water phase diagram and the critical points under pressure

There is no more first-order transition at a critical point. By applying equations (2) and (3) and assuming $\Delta\varepsilon = 0$, and $P_1 - P_2 = 0$, Lines 1 and 2 at a critical point in Figure 5 are shifted by $P = P_1 = P_2$ under pressure. In Figure 10, the LLPT line, $\theta_{LL} = -0.167$ ($T_{LL} = 0.832 \times T_m$), extends from P=-0.5000 to P=0.8505. A reduced temperature is used because it reduces the figure number and it may apply to the melting temperature of any ice phase.

The critical points are determined assuming that glass Phase 3 continues to exist as a liquid phase when heated above $\theta_g$. A complementary volume change, corresponding to the HDA-VHDA transformation is observed at 125 K under higher pressures (approximately p=0.95GPa) [11, Figure 1]. and is frozen after decompression at 77 K and equal to 0.0855 m$^3$g$^{-1}$. This transition is due to superheating of Phase 3 which disappears at the second homogeneous nucleation temperature when $\theta_{n+} = \Delta\varepsilon_{lg} = 1.302 \times T_m$ and it is accompanied by an enthalpy increase equal to $0.302 \times \Delta H_m = 1812$ J.g.$^{-1}$mole$^{-1}$ in the strong liquid and a volume change $\delta V=1812/18/p=10^{-6}m^3g.^{-1}$ (which agrees with [11]). Equation (17) gives the value of $\theta_{n+}$ for all glass phases as a function of their initial enthalpy coefficient [1]:

$$\theta_{n+} = \frac{-1 + (1 + 4(\varepsilon_{ls0} - \varepsilon_{gs0} + \Delta P)(\varepsilon_{ls0}\theta_{0m}^{-2} - \varepsilon_{gs0}\theta_{0g}^{-2}))^{1/2}}{2(\varepsilon_{ls0}\theta_{0m}^{-2} - \varepsilon_{gs0}\theta_{0g}^{-2})} \qquad (17)$$

where $\varepsilon_{ls0}=1.14286$, $\varepsilon_{gs0}=0.66667$, $\theta_{0g}=-1$, $\theta_{0m}=-2/3$ and $\Delta P=-0.3704$ for LDA.

The melting temperature ($T_m$) of ice, Ih, under 0.95GPa is deduced to be equal to $125/1.302=96$K which is in agreement with Mishima's measurements [9]. The existence of a melting temperature above $T_m$ due to Phase 3 superheating suggests that any liquid Phase 3 is "ordered" above $T_g$ and $T_m$. The existence of an "ordered liquid" state has already been suggested to exist above $T_g$ and above $T_m$ in $Zr_{41.2}Ti_{13.8}Cu_{12.5}Ni_{10}Be_{22.5}$ [63,64]. The temperature $T_{n+}$ is observed from 1090 to 1150 K and equation (17) predicts 1116K [1]. An other glass-forming melt ($Zr_{58.5}Cu_{15.6}Ni_{12.8}Al_{10.3}Nb_{2.8}$) is ordered below 850K. Its glass transition temperature is 700 K and $T_m=1125$ K. Using equation (12) and a=1 because $\Delta C_p(T_g)=1.5\times\Delta H_m/T_m$, $\theta_g=-0.378$, $\varepsilon_{ls0}=\theta_g+2=1.622$ and $1.5\times\theta_1=\theta_g$; the homogeneous nucleation temperature $T_1$ in Phase 1 is calculated to be equal to 842 K. A specific heat peak is observed in Figure 11 at this temperature which shows that liquid Phase 1 is ordered below its homogeneous nucleation temperature [65]. The reference of the specific heat data is [66].

There is no more first-order transition above P=0.8505 because the homogeneous nucleation temperatures of strong Phase 1 and Phase 2, calculated using equations (5) and (7), reappear above $\theta_{LL}$ as shown in Figure 10.

Another point without first-order transition occurs for $P_1=0.006$ because the sum of latent heats of Phase 1 and Phase 3 at $\theta_{LL}$ is nearly equal to zero, as seen in Figure 5. These two points occur for $p_2/\rho_2=283.5$ and $p_1/\rho_1=2$ where $\rho$ is the density in Kg.m$^{-3}$ and p the pressure in Pascal. The pressure $p_1$ is equal to 18.3 MPa, where $\rho_1$ =915 and $p_2$ =0.31 GPa where $\rho_2=1093$ which is in rough agreement with other calculations [23,27,29,30] and with density measurements under pressure [67-69]. Liquid Phase 3 exists above $T_g$ and this explains the presence of a point at a low pressure equal to approximately 18.3MPa where the first-order transition disappears without being the end of the first-order transition line. The corresponding pressure slightly depends on the initial choice of $T_g$ but, it is equal to approximately P=0 for $T_g=135$ K instead of 136.6 K.

The specific heat increase at zero pressure below $T_m$ proves that the LLPT is always present for P<0.006 and exists at negative pressures down to P=-0.500. Phase 3 disappears with the glass transition for P=-0.500 when Line 1 and Line 2 in Figure 5 are shifted by -0.500. This third point is critical because it corresponds to the other extremity of the first-order transition line and confirms that the stability limit of the two metastable water phases occurs for p=-175 MPa assuming a density $\rho=950$ Kg.m$^{-3}$ [70]. This last critical point also corresponds in Figure 5 to the highest exothermic enthalpy under pressure and then, to the maximum density at negative pressure [70,71]. When assuming zero pressure, equations (2), (3), and (10) indicate that the first-order transformation during heating could be endothermic between P=0.006 and P=0.8505 and exothermic for -0.5<P<0.006 adding Phase 1 and Phase 3 contributions to the latent heat (see Figure 5).

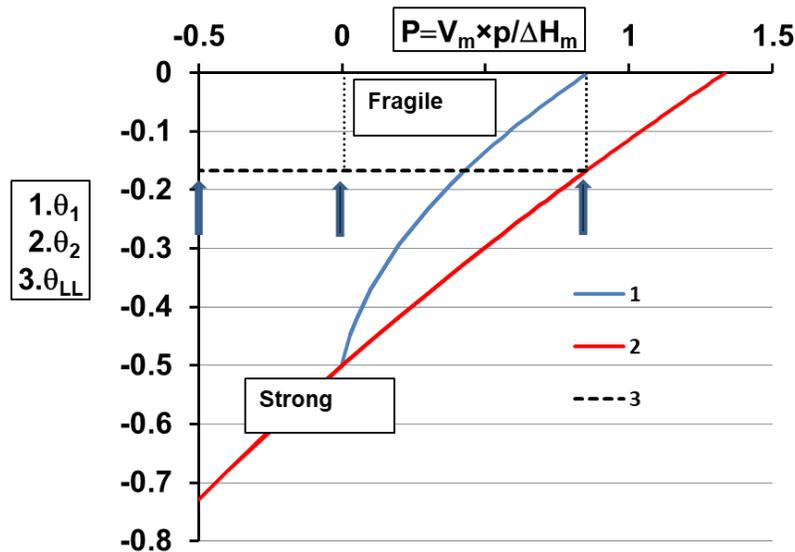

**Figure 10** : **Phase diagram of supercooled water at pressure p**. *Curves numbered from 1 to 3: 1- Homogeneous nucleation temperature of strong Phase 1 versus the enthalpy coefficient P induced by the pressure p; 2- Homogeneous nucleation temperature of strong Phase 2 versus the enthalpy coefficient P induced by the pressure p; 3- LLPT line separating the fragile liquid phase from the strong one at $\theta_{LL}$=- 0.16715; first critical point:P=-0.5, p≅175 MPa; second point: P=0.006, p≅18.3 MPa; third point at the end of first-order line, P=0.8505, p≅0.31 GPa.*

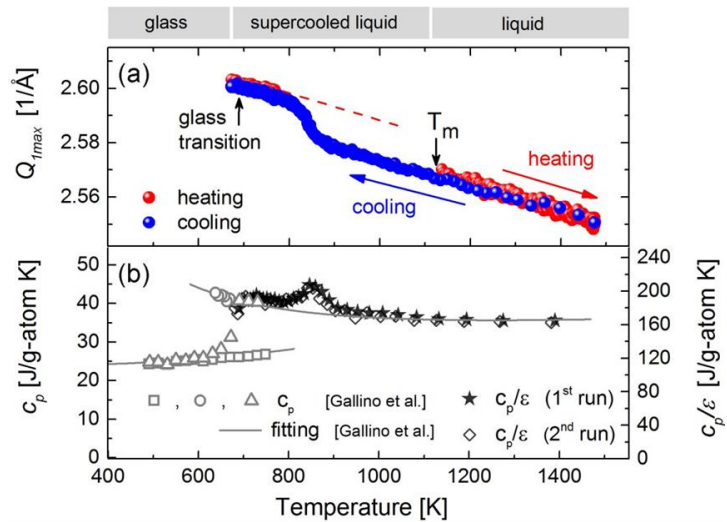

**Figure 11**: *$Zr_{58.5}Cu_{15.6}Ni_{12.8}Al_{10.3}Nb_{2.8}$. Reprinted from [65] with APS permission."(a) Position of the first maximum of S $*(Q)$,Q1max, during heating of an initially glassy sample and subsequent cooling down to the glassy state. (b) Ratio of specific heat capacity to total hemispherical emissivity $c_p/\varepsilon$ calculated from*

*the temperature profile measured in ESL during cooling to the glassy state in comparison with the measured calorimetric cp data of Gallino et al. [66]".*

**Conclusions**:

The thermodynamic parameters of two water phases (Phase 1 and Phase 2), separated by an enthalpy difference depending on $\theta^2=(T-T_m)^2/T_m^2$, have been determined only knowing the formation temperature of a strong glass in Phase 3 where $T_g$=136.6 K, the first-order LLPT was at -45.6°C in confined water under pressure, the compressibility minimum was +45.6 °C, the ice melting heat was $\Delta H_m$=6000 J.mole$^{-1}$, and the melting temperature was 273.1 K.

The LDA phase of strong glass contains an enthalpy excess below $T_g$=136.6K, resulting from quenching. Consequently (below $T_g$=136.6K) a sharp, exothermic latent heat is observed through relaxation heating after total decompression at 77K and it can be predicted to occur at temperatures ($T_{sg}$) in agreement with experimental results. The maximum relaxed enthalpy cannot be higher than its value at the Kauzmann temperature even if the frozen enthalpy is equal to -0.3704×$\Delta H_m$ at $\theta$=-2/3 without entropy constraint. The enthalpy excess present in the bulk glass at the transformation temperature, leads to partially-amorphous ice instead of fully-relaxed ultrastable glass because the LDA density is too close to that of ice. LDA exists under various forms because any ice having a fusion entropy smaller than 22 J.K.$^{-1}$mole$^{-1}$ is crystallized at a temperature equal or larger that its own Kauzmann temperature.

A sharp volume increase from HDA to LDA has been measured at 115.5 K. This transformation temperature corresponds to the Kauzmann temperature $T_K$=$T_{sg}$≅115.5 K. This is the first observation of the Kauzmann temperature of a glass.

Supercooled water is a fragile liquid above a liquid-to-liquid phase transition (LLPT) at $T_{LL}$=0.833×$T_m$ and it is transformed into a strong liquid below $T_{LL}$. The first-order character of LLPT disappears for three pressures equal to approximately -175MPa, 18.3MPa, and 310 MPa. Glass Phase 3 disappears for p=-175MPa because there is no more glass transition below this negative pressure. For the first time, it has been shown that glass Phase 3 is transformed into a new liquid phase above $T_g$ and that the two liquids separated below $T_{LL}$ are liquid Phase 1 and liquid Phase 3 for -0.175 GPa<p<+0.31 GPa. Along the LLPT line, the first-order transition could be exothermic by heating from -175MPa to 18.3MPa and endothermic from 18.3 MPa to 310 MPa.

Double glass transitions are expected under pressure at $T_{sg}$ and $T_g$ when the glass enthalpy is still enhanced by an enthalpy excess. The glass transition occurs at $T_g$=0.5×$T_m$ at low pressure (p<0.31 GPa) and at 0.632×$T_m$ for HDA under high pressure (0.31<p<0.6 GPa). The first-order LDA-HDA phase transitions at $T_{sg}$ under pressure can be predicted, leading to constant volume and enthalpy changes. The predictions correspond to the formation of a new glass Phase 3 with an enthalpy increase equal to the maximum frozen enthalpy (0.3704×$\Delta H_m$) available at T=91K ($\theta$=-2/3). This enthalpy change is no longer limited by its value at the Kauzmann temperature because this glass phase is induced by a first-order transition; it has an entropy maximum reduction at $T_{sg}$ which is equal to the available entropy below its own glass transition. The entropy and enthalpy changes at $T_{sg}$ are expected to be recovered in the "no man's land" at this new $T_g$.

Phase 3 does not disappear but continues to exist as an "ordered" liquid phase above $T_g$. Ordered liquid Phase 3 is superheated above $T_m$ and disappears at the liquid homogeneous nucleation temperature $T_{n+}$. This transition is accompanied by a sharp volume decrease under pressure. VHDA is identified as being formed due to the melting of this ordered liquid Phase 3.

All these theoretical findings, using classical nucleation theory completed by an enthalpy difference between two liquids, are fully compatible with the experimental results (without introducing any complementary parameter to ensure the fit). The existence of an ordered liquid above $T_g$ and $T_m$ has been suggested by other authors can now be confirmed, without knowing the nature of its microscopic order at the atomic scale. Ordered phases have to exist in all glass-forming melts giving rise to various glass phases. This work was based on the prediction of homogeneous nucleation temperatures $T_{n-}$ of various liquid and glass phases and it is suggested that they are the formation temperatures of new ordered phases of superclusters followed, after subsequent cooling, by the percolation threshold of dynamical fractal structures above $T_g$. All these new "ordered phases" still have superheating temperatures $T_{n+}$ at the second homogeneous nucleation temperature $T_{n+}$ above $T_m$.

**Acknowledgments**: The author thanks M.I. Ojovan for discussions about the thermodynamic origin of the glass transition. Several tentatives to find the temperature $T_{n+}$ of bulk water using DSC measurements by Gael Moiroux and J.L. Garden in the Neel Institute have failed. Measurements of the temperature gradient along a vertical axis in a recipient filled with water have been made by J. Blanchart. Criticisms of successive anonymous reviewers of Chemical Physics Letters and Chemical Physics have contributed to clarify the presentation of this work.